\documentclass[twocolumn,showpacs,preprintnumbers]{revtex4-1}

\usepackage{graphicx}
\usepackage{color}
\usepackage{longtable}

\begin{document}

\title{Possible Coexistence of Superconductivity and Magnetism in Intermetallic
NiBi$_3$}
\author{Esmeralda Lizet Martinez Pi\~{n}eiro, Brenda Lizette Ruiz Herrera, and
Roberto Escudero}
\email[correspondence author: ]{escu@servidor.unam.mx}
\affiliation{Instituto de Investigaciones en Materiales, Universidad Nacional Aut\'{o}noma de M\'{e}xico. 
A. Postal 70-360. M\'{e}xico, D.F. 04510 M\'EXICO.}
\author{Lauro Bucio}
\affiliation{Instituto de F\'isica, Universidad Nacional Aut\'{o}noma de M\'{e}xico. 
M\'{e}xico, D.F. 04510 M\'EXICO.}

\date{\today}

\begin{abstract}
NiBi$_{3}$ polycrystals were synthesized via a solid state method. X-ray
diffraction analysis shows that the main phase present in the sample
corresponds to NiBi$_{3}$ in a weight fraction of 96.82 $\%$ according to
the refinement of the crystalline structure. SEM - EDS and XPS analysis reveal a homogeneous
composition of NiBi$_{3}$, without Ni traces. The powder superconducting
samples were studied by performing magnetic measurements. The superconducting
transition temperature and critical magnetic fields were determinated as T$_{C}$ =
4.05 K, H$_{C1}$ = 110 Oe and H$_{C2}$ = 3,620 Oe. The superconducting parameters
 were $\xi_{GL}$= 301.5 $\mathring{A}$, $
\lambda_{GL}$= 1549 $\mathring{A}$, and $\kappa$= 5.136. Isothermal
measurements below the transition temperature show an anomalous behavior.
Above the superconducting transition the compound presents ferromagnetic
characteristics up to 750 K, well above the Ni Curie temperature.
\end{abstract}

\pacs{Intermetallic Alloys; Superconductivity; Ferromagnetism;}
\maketitle

\section{Introduction}

The coexistence of superconductivity and magnetism is a phenomena of great
interest in the scientific community. In 1957, Ginzburg \cite{ginzburg} considered that the
coexistence could exist if the critical field were longer that the
induction created by the magnetization. Two years before the antagonic
nature of superconductivity and magnetism was confirmed when Matthias'
experiments \cite{mattias superc} showed that the superconductivity in
lanthanum was destroyed by a small concentration of magnetic impurities.
This was explained as in conventional \textit{s}-wave superconductors, local magnetic
moments break up spin singlet cooper pairs and hence strongly suppress
superconductivity, an effect known as magnetic pair-breaking. Because of the
pair-breaking effect, in most superconductors the presence of only a 1\%
level of magnetic impurity can result in the almost complete destruction of
the superconducting behavior.

The discovery of rare earth ternary and actinide compounds was the first
opportunity to study the interaction between the magnetic moments of
\textit{f}-electrons and superconducting electrons in a very high density of local
moments \cite{kakani,Sinha,Khan}. These compounds presented new exotic phenomena
associated with the long-range order of local magnetic moments, such as
reentrant superconductivity \cite{ishikawa}, coexistence of superconductivity and
ferro-antiferromagnetism \cite{hamakar}, magnetic field inducing  superconductivity \cite{maple} and heavy fermion superconductivity \cite{steglich}.

NiBi$_{3}$ is an intermetallic alloy with orthorrombic structure, CaLiSi$_{2}$-type and space group Pnma \cite{Vassilev,Park}. In this structure, bismuth atoms form an octahedral array, while nickel atoms form  part of linear chains. NiBi$_{3}$ is a superconducting material with a
critical temperature about 4.05 K \cite{alek}, and is the object of study of this research.

Some initial studies has been done on the superconducting properties of NiBi$_{3}$. For instances Fujimori, et al. \cite{fujimori} have presented a study
related to the superconducting and normal properties; they studied the
resistivity, heat capacity, upper critical magnetic field, in
polycrystals and needle like single crystals. 

Among the electronic characteristics of this intermetallic alloy is that it presents a large
phonon resistivity due to predominant coupling of electrons by bismuth
vibrations via the Ni vibrations. In this study we are mainly concerned with the magnetic-superconducting interacting effects, as we will discussed in the rest of the paper.

\section{Experimental Details}

Several samples were studied. The preparation method for the most pure obtained sample was the following: the NiBi$_{3}$ 
composition was  prepared by solid state method using Bi pieces
(Aldrich 99.999 \%) and Ni powder (Strem Chemicals 99.9 \%) in evacuated
quartz tubes. The samples were melted in a resistance furnace at ${1000}^{\circ}$ C for seven days. The characterization 
was made by X-ray powder diffraction (XRD) (Bruker AXS D8 Advance) using Cu K$\alpha$ radiation, scanning electron microscopy
 (SEM - EDS) (Leica-Cambridge), and X -ray photoelectron   spectroscopy (XPS) (Microtech Multilab ESCA2000) using Al K$\alpha$
 radiation 1453.6 eV. The XPS spectra was obtained in the constant pass energy mode (CAE) E$_0$ = 50 and 20 eV for survey and high resolution respectively. The sample was etched for 20 minutes with Ar$^+$ at 3.5 kV during 20 minutes 
at 0.12 $\mu$A mm$^{-2}$. The peak binding energy (BE) positions were referenced to Au 4f$_{7/2}$ at 84.00 eV and Ag 3d$_{5/2}$ at 367.30 eV having a FWHM of 1.02 eV. Magnetization measurements and determination of
the superconducting properties were performed using a Quantum Design (QD)
superconducting quantum interference device (SQUID) MPMS system. We determined the transition temperature
with magnetization versus temperature M(T) measurements  with a small magnetic field about 10 Oe.
 Two normal distinct measuring modes were used, zero field cooling
(ZFC) and field cooling (FC). Critical magnetic fields were determined by
performing isothermal measurements M(H) at different temperatures, from 2 to 4K. 
We also performed isothermal measurements at different
magnetic intensities at higher temperatures to observe possible magnetic behavior above the transition temperature and at much higher temperatures. In order to determine the Curie temperature and to discard the possibility of Ni as an impurity in the compound, we performed studies to observe in the isothermal curves, M-H,  the existance of hysteresis, thus the coercitive field at different temperatures, and well above the Ni and NiBi Curie temperatures using a Quatum Design oven installed in the MPMS.   

\section{Results and Discussion}

\subsection{Structural characterization}

The structural characterization determined by powder XRD and Rietveld refinement are shown in Fig. 1. The polycristalline  phases were identified by
comparison with X-ray patterns in the Inorganic Crystal Structure Database (ICSD) 2010. All the peaks
correspond to the  NiBi$_{3}$ phase (ICSD 58821), except the weak peaks related to impurities
of Bi (ICSD 64703) and NiBi (ICSD 107493), in a proportion less than 1 and 2 \% respectively. According
to the results of the refinement, no Ni impurities were detected.

 \bigskip

\begin{figure}[btp]
\begin{center}
\includegraphics[scale=0.35]{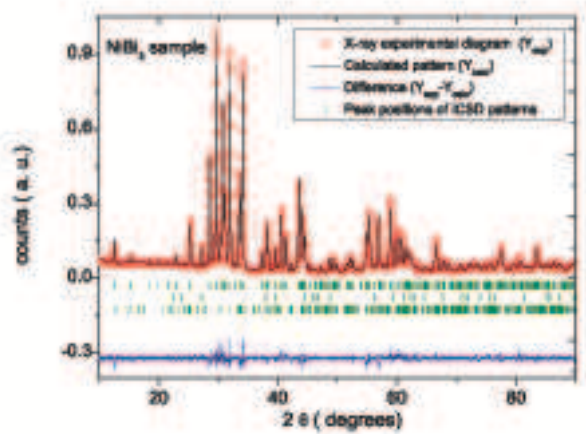}
\end{center}
\caption{(Color online) Rietveld refinement of NiBi$_3$ sample. Dots are the experimental data, the continuous line superposed to the dots is the calculated pattern. 
At the bottom of the diagram, the difference between the experimental and calculated points is shown. Vertical marks are displayed in three rows corresponding from top to bottom to the Bragg positions for the phase NiBi$_3$, Bi, and NiBi respectively.}
\label{fig1}
\end{figure}

Refinement of the crystalline structure was performed using a Rietveld Fullprof program.
Table 1  contains the NiBi$_{3}$ structural 
parameters, corresponding to orthorhombic space group Pnma (62). In figure 2 is shown the crystalline structure. 

\bigskip 
\begin{table}[tbp]
\caption{Crystallographic data for NiBi$_{3}$ obtained by Rietveld
refinement starting with the structural parameters reported by Fjellvag and Furuseth \cite{FJ} with symmetry described by the orthorhombic space group Pnma. The standard deviations are written between parentheses. }
\label{table 1}
\begin{tabular}{cp{2.5 cm}p{2cm}p{2cm}p{2cm}}
\hline
& Rp(\%)  & Rwp(\%) & Re(\%) & $\chi^{2}$  \cite{young} \\ \hline
  & 14.0 & 15.3 & 11.1 & 1.886  \\ \hline
& Parameters ($\mathring{A}$)  & a = 8.879(1)  & b =  4.0998(7)  &  c = 11.483(2) \\ 
& & & & \\ 
& Volume ($\mathring{A}^{3}$) & 417.7(2) & &\\ \hline
& site & x & y & z \\ \hline
& Bi 1 & 0.298(3) & 0.25 & 0.890(2) \\ 
& & & & \\ 
& Bi 2 & 0.382(4) & 0.25 & 0.594(2) \\ 
& & & & \\ 
& Bi 3 & 0.409(3) & 0.25 & 0.180(2) \\ 
& & & & \\ 
& Ni 1 & 0.09(1) & 0.25 & 0.520(6) \\ 
\hline
\end{tabular}
\end{table}

\bigskip

\begin{figure}[btp]
\begin{center}
\includegraphics[scale=0.3]{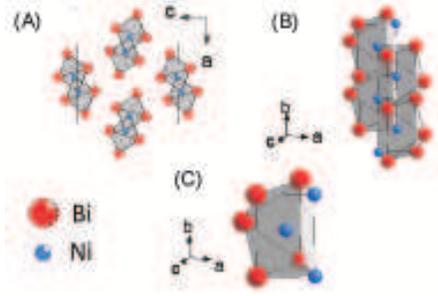}
\end{center}
\caption{(Color online) NiBi$_3$ crystalline structure. (A) Unit cell; (B) Rods of prisms oriented along the b axis; 
(C) Ni atoms with capped trigonal prismatic coordination and strong bonds Ni-Bi and Ni-Ni. }
\label{fig2}
\end{figure}

\bigskip

Yoshida et al.\cite{Yoshida} reported magnetic properties of NiBi. Unlike their samples  with important amounts of nickel impurities,
our NiBi$_{3}$ samples are completely free of nickel, according to the X-
ray diffraction analysis, refinement, and XPS studies as we will show below.

Figure 3 shows SEM - EDS images of NiBi$_{3}$. A
comparison between two of our samples ($a$, $b$) with different reaction times; with
three and seven days at $1000^\circ$ C are presented, respectively. Sample $a$ shows the presence of two phases, NiBi and NiBi$_{3}
$, differentiated by the color and borders clearly defined. Sample $b$ shows
a homogeneous composition of NiBi$_{3}$. Impurities of NiBi and Bi were not detected by EDS.

\bigskip

\begin{figure}[btp]
\begin{center}
\includegraphics[scale=1.0]{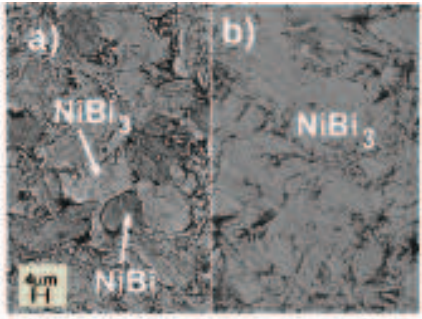}
\end{center}
\caption{SEM Images of NiBi$_{3}$. a) NiBi$_{3}$ after heating at 1000 ${
{}^\circ}C$ during 3 days, b) and during 7 days.}
\label{fig3}
\end{figure}

In order to discard nickel impurities, XPS analysis was performed in our samples. Figure 4a shows the survey spectra for NiBi$_3$. It was observed that all peaks correspond to Ni and Bi, any  trace of other elements were found. Figure 4b shows the XPS spectra at low binding energies for the NiBi$_3$ sample and its comparison with bismuth and nickel metal references. The analysis of the sample in the valence region shows that the $3d$ peak of nickel spectrum displays a  increasing width and movement to higher energies with respect to nickel metal. This effect is attributed to changes in the environment of the nickel atoms when forming part of the alloy, confirming that nickel atoms are part of the NiBi$_3$ compound and not as an impurity.

\bigskip

\begin{figure}[btp]
\begin{center}
\includegraphics[scale=0.3]{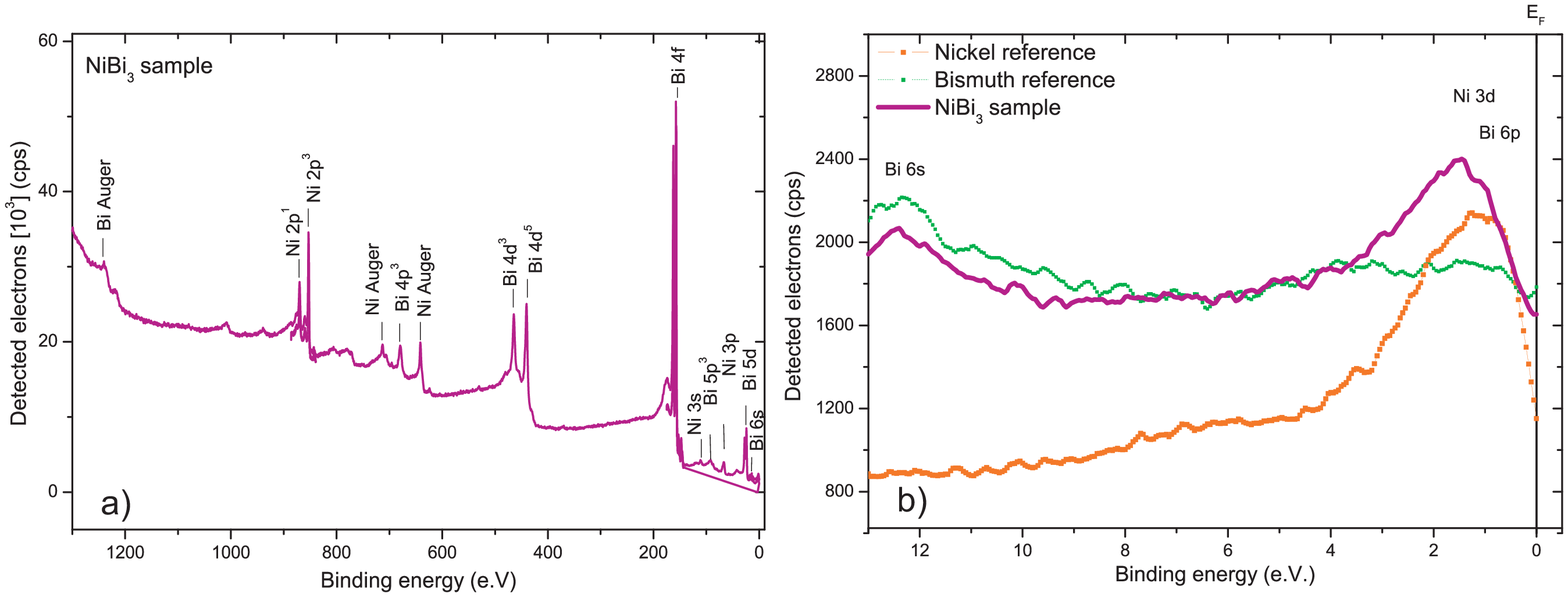}
\end{center}
\caption{(Color online) XPS analysis of NiBi$_{3}$. a). XPS survey spectra  for NiBi$_3$. b). XPS valence band spectra for Ni, Bi, and NiBi$_3$.This XPS analysis shows that the compound is a phase without Ni impurities.}
\label{fig4}
\end{figure}

\subsection{Superconductivity}

Magnetization measurements M(T) performed in ZFC and FC modes were performed
in order to determine the amount of superconducting fraction.  The Meissner fraction of NiBi$_{3}$
sample was determined to be about 54.1\% and was calculated a 2 K related to 
the maximum value $-4\pi \chi =-4\pi \rho M/mB$, where $\rho $ is the density of the
material equal to 10.884 g/cm$^{3}$, $M$ is the magnetization in
emu, $m$ is the mass in g, and $B$ is the applied magnetic field \cite{porc}. The superconducting transition
temperature T$_{C}$ was 4.05 K and is defined as the point in which there
is a drop in the susceptibility in the FC measurement. These
results are shown in the Fig. 5.

\bigskip

\begin{figure}[btp]
\begin{center}
\includegraphics[scale=0.25]{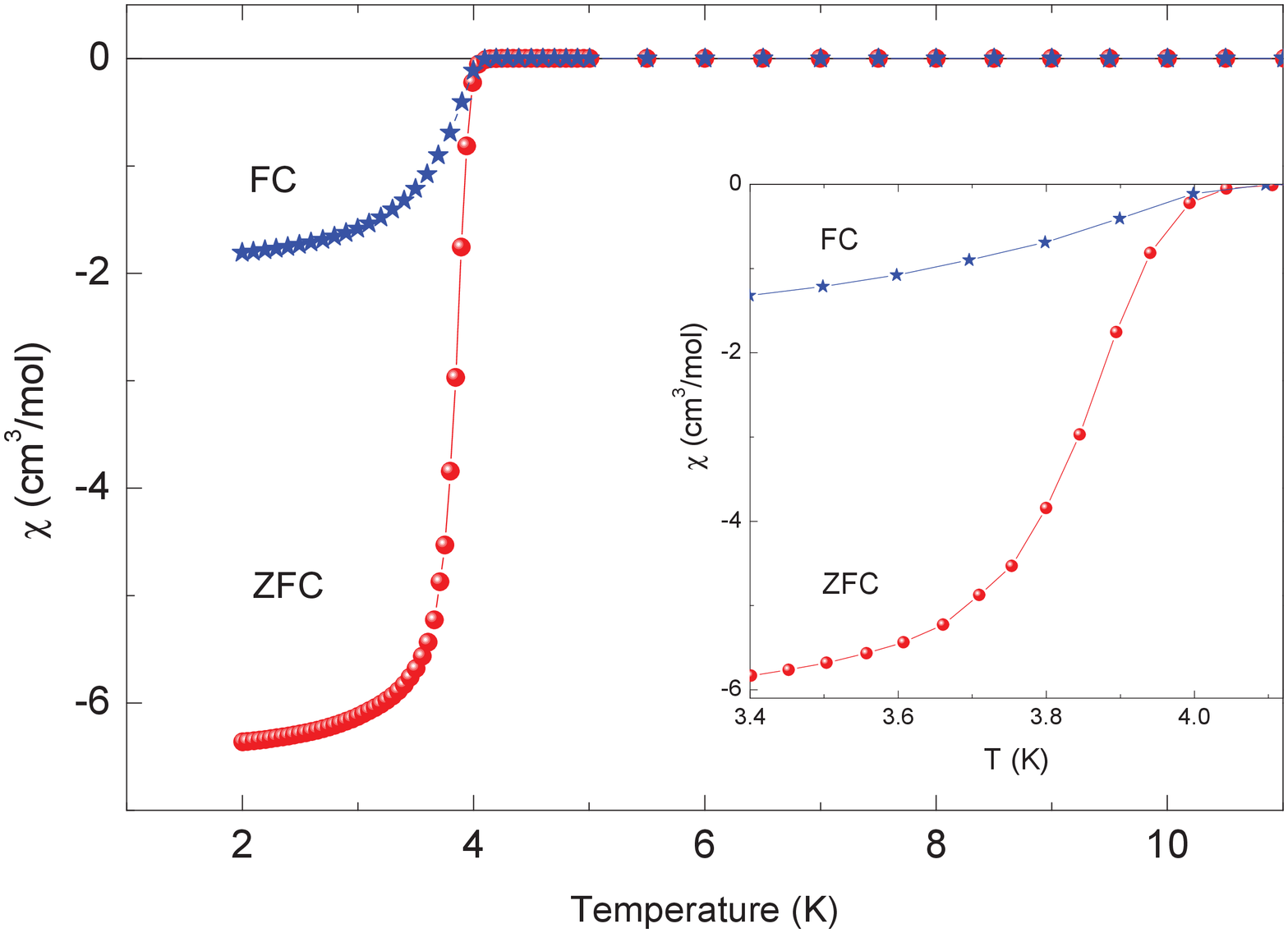}
\end{center}
\caption{(Color online) Shielding and Meissner fractions measured at 10 Oe. The
inset displays the measurements close to the transition temperture. The onset of the transition is about 4.05 K}
\label{fig5}
\end{figure}

\subsection{Magnetic measurements.}

\bigskip\ 

Magnetization measurements as a function of applied magnetic field M(H),
were used to calculate the superconducting parameters. The critical magnetic fields H$_{C1}(0)$ and H$_{C2}(0)$ 
were calculated from the experimental data and a linear fit using 
the expression H$_{Ci}(0)=-0.693$$T_{C}$$(dH_{Ci}(T)/dT\mid_{T=T_{C}})$
near T$_{C}$, where $dH_{Ci}(T)/dT\mid_{T=T_{C}}$ corresponds to the
slope of the linear fit \cite{parametros}. Figure 6 shows the critical
magnetic fields. 

\bigskip

\begin{figure}[btp]
\begin{center}
\includegraphics[scale=0.15]{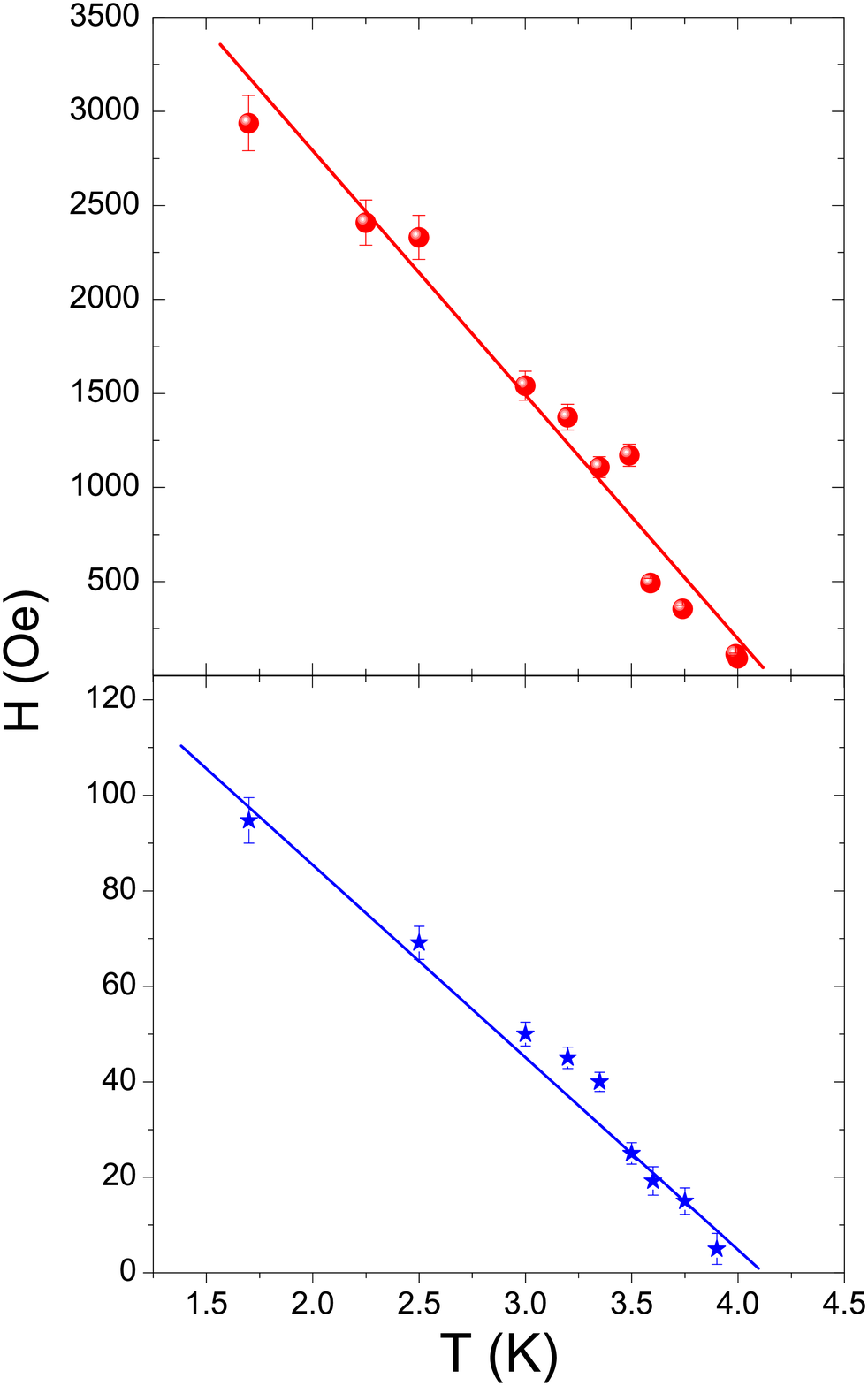}
\end{center}
\caption{(Color online) Critical fields $H_{C1}$, and $H_{C2}$ as function of temperature. Critical fields were measured using isothermal magnetic
 curves, data were fit with the expressions mentioned in the main text.}
\label{fig6}
\end{figure}

The Ginzburg-Landau (GL) parameters; coherence
length $\xi _{GL}$, penetration length $\lambda_{GL}$, $\kappa$, and the
thermodynamical critical field H$_{C}(0)$, were estimated with  the 
equations: $H_{C2}(0)$=$\phi_{0}/2\pi\xi_{GL}^{2}$, $H_{C2}(0)/H_{C1}(0)$=2$\kappa^{2}/\ln\kappa$, $\kappa$=$\lambda_{GL}/\xi_{GL}$, and $H_{C}(0)$=$\phi_{0}/(2\sqrt{2}\pi\xi_{GL}\lambda_{GL})$ where $\phi{_0}$ is the quantum flux. Also we may use $H_{C1}H_{C2}=H{_C}\ln\kappa$. 
The superconducting parameters are presented in table II and are
similar to the obtained by Fujimori, et al. \cite{fujimori}.

\begin{figure}[btp]
\begin{center}
\includegraphics[scale=0.4]{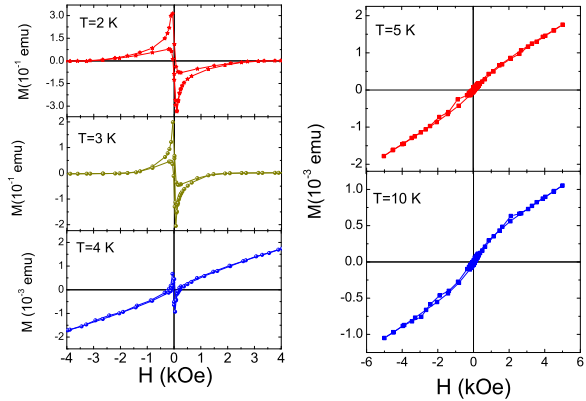}
\end{center}
\caption{(Color online) Isothermal magnetic measurements M-H in the superconducting region of NiBi$_3$, note the anomalous characteristics of data, which is the effect of ferromagnetism in the sample.}
\label{fig7}
\end{figure}

\bigskip 
\begin{table}[tbp]
\caption{Superconducting parameters of NiBi$_3$ compound}
\begin{tabular}{p{1cm}p{3cm}crlp{2cm}}
\hline
& & & & & \\
&\textbf{Property} & \multicolumn{3}{c}{\textbf{Value}} &  \\ \hline
& & & & & \\
& T$_{C}$ & & 4.05 &$\pm$ 0.1 & K \\ \bigskip
& & & & & \\ 
& $dH_{C1}(T)/dT\mid_{T=T_{C}}$ & & 40 &$\pm$ 3 & Oe/K \\ 
& & & & & \\
& $dH_{C2}(T)/dT\mid_{T=T_{C}}$ & & 1290 &$\pm$ 80 & Oe/K \\ 
& & & & & \\
& H$_{C1}(0)$ & & 110 &$\pm$ 8 & Oe \\ 
& & & & & \\
& H$_{C2}(0)$ & & 3620 &$\pm$ 300 & Oe \\ 
& & & & & \\
& $\xi_{GL}$ & & 302 &$\pm$ 10 & ${\mathring{A}}$ \\ 
& & & & & \\
& $\lambda_{GL}$ & & 1549 &$\pm$ 100& ${\mathring{A}}$ \\ 
& & & & & \\
& $\kappa$ & & 5.1 &$\pm$ 0.2 & \\ 
& & & & & \\
& H$_{C}(0)$ & & 490 &$\pm$ 30 & Oe \\ \hline
& & & & & \\
\end{tabular}
\end{table}

\begin{figure}[btp]
\begin{center}
\includegraphics[scale=0.4]{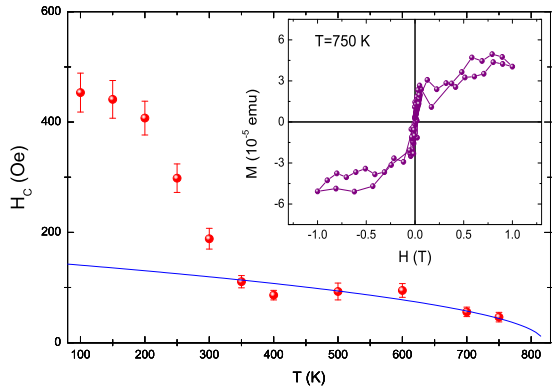}
\end{center}
\caption{(Color online) Coercive field extracted from  M-H measurements from 100 to  750 K. Clearly the ferromagnetic behavior is observed, and also the  coercive field. Accordingly the ferromagnetic transition is above 800 K, and consequently the only magnetic contribution is of NiBi$_3$. The inset shows the isothermal measurements at 750 K. At that temperature the ferromagnetic characteristic is still clearly observed  at low fields.}
\label{fig8}
\end{figure}
\bigskip 

\subsection{Magnetic measurements below and above T$_C$}

As was mentioned before different magnetic measurements were performed to have a better insight about the electronic properties of this
alloy. Below the transition temperature M(H) measurements were used to  determinate  the critical fields in the usual manner; thus fitting a straight line to the M(H) curve at different temperatures. Below T$_C$ at the temperatures of 2, 3, and  4 K, Fig. 7 displays the competition between the diamagnetic characteristic of the superconducting state and the ferromagnetism of the compound. At low field M(H) curves looks normal, so H$_{C1}$ can be determined. However as soon as the maximum diamagnetism is reached, a fast decrease of the diamagnetic contribution, $-M$ is observed. This decreasing  characteristic changes more  rapidly that in normal superconductors (where the magnetism is absent). This anomalous behavior is the indication of two competing proceses: superconductivity and ferromagnetism. At high temperature above the transition temperature, where the diamagnetic characteristic  disappears, M(H) shows a typically ferromagnetic characteristic; thus a tendency to saturation of the magnetization at high fields and a coercive field, h$_C$, at the central part of the curve. 

M(H) mesurements were performed in order to determinate the coercive field h$_C$. We used a medium field model above the superconducting temperature  given by:  
$h{_C}$(T) = h${_C}$(0)[1-(T/$\tau$${_C}$)${^{1/2}}$], in this  equation $\tau$${_C}$ is the Curie temperature. In Fig. 8 we present the variations of h${_C}$ up to 750 K. The fitting line is the result of the above equation for h${_C}$. At this  high temperature we observed in the inset the  M(H) data obtained  at 750 K. This last measurement is above the Curie temperature of Ni 355${{}^\circ}C$ ( or 627 K \cite{kittel}). The measurement performed at 750 K clearly indicates that NiBi$_3$ is the only magnetic contributing material. In the Yoshida's 
studies they  observed that NiBi is magnetic at a maximum about 640 K.  Our measurements show that at  750 K  the coercive field is small and about 70 Oe.

\section{Conclusions}

We found that NiBi$_{3}$ is a type II superconducting material  in which
ferromagnetism and superconductivity coexist and presents an interesting
interplay. This is the first time that such coexistance is demonstrated in a
clear experimental manner. The ferromagnetic transition is persistent at
very high temperature about 750 K, well above the Curie temperature of Ni metal and NiBi compound.

\begin{acknowledgments}
We thank F. Silvar for Helium provisions, Lazaro Huerta for the XPS
measurements, Francisco M. Ascencio for his collaboration in the sample
preparation.
\end{acknowledgments}

\end{document}